\documentclass[osajnl,preprint,showpacs,superscriptaddress,12pt]{revtex4-1} %% use 12pt for preprint option
\usepackage{amsmath,amssymb,graphicx}
\begin{document}

\title{Generation of~multi-terawatt vortex laser beams}

\author{Craig~Ament} 
\author{Lee~Johnson}
\affiliation{College of Optical Sciences, The University of Arizona, Tucson, Arizona 85721, USA}
\author{Andreas~Schmitt-Sody}
\author{Adrian~Lucero}
\affiliation{Air Force Research Labs, Kirtland Air Force Base, Albuquerque, New Mexico 87117, USA}
\author{Thomas~Milster}
\author{Pavel~Polynkin}\email{Corresponding author: ppolynkin@optics.arizona.edu}
\affiliation{College of Optical Sciences, The University of Arizona, Tucson, Arizona 85721, USA}

\begin{abstract}
We report the fabrication of large-area phase masks on thin fused-silica substrates that are suitable for shaping
multi-terawatt femtosecond laser beams. We apply these phase masks for the generation of intense femtosecond 
optical vortices. We further quantify distortions of the vortex beam patterns that result from several common types 
of mask defects.
\end{abstract}

%\ocis{(320.7110)  Ultrafast nonlinear optics; (140.3300) Laser beam shaping; (070.6120) Spatial light modulators}

\maketitle

\section{Introduction}

Laser beam shaping is a concept as old as the laser itself and is essential in such diverse applications of the laser as
material processing, lithography and biology. 
Recently, laser beam shaping has been applied to the studies of self-action effects of ultra-intense femtosecond 
laser pulses in air. Phase and amplitude perturbations across the shaped beam profile act as 
nucleation sites that determine the placement of intense filaments resulting from self-focusing of the beam
\cite{filament1}. Various complex beam shapes have been applied to filamentation studies in air including
fundamental Bessel beams \cite{Bessel1,Bessel2,Bessel3}, Airy beams \cite{Airy} and optical vortices \cite{fsVortex,vortexPRL}.
Propagation dynamics of intense optical vortices in air is particularly interesting. 
It has been argued that the bottle-like distributions of dilute plasma filaments that are created in air through
self-focusing of such beams can act as extended conduits for microwave radiation \cite{MW1,MW2}.

Airy and vortex laser beam patterns investigated in \cite{Airy,vortexPRL} have been produced through the modulation
of initially smooth input beam profiles with specially designed transmissive phase masks.
Those masks have been made by profiling thin polyamide coatings atop fused silica substrates. They have an estimated damage
threshold of the order of 10\,mJ/cm$^{2}$, for laser pulses with about 40\,fs duration. For the centimeter-scale 
input laser beam diameter, the above damage threshold corresponds to the maximum pulse energy of the order of 
10\,mJ. Such pulse energies are adequate for laboratory-scale investigations with the meter-scale extent of the
generated plasma filaments. For real outdoor applications that require filament lengths of tens to hundreds of meters,
the input pulse energy has to be scaled to hundreds of mJ and beyond. In that case polyimide phase masks
become inadequate as they are easily damaged by the laser beam on a single laser shot. The application of either
purely reflective phase masks or transmissive masks made of an optical material with high damage threshold 
is necessary. 

In this paper, we report the fabrication of phase masks with area of up to 50\,mm\,$\times$\,50\,mm
made on 0.5\,mm-thick fused silica substrates. We have experimentally confirmed that these masks withstand
an extended exposure to femtosecond laser beams with energy per pulse of up to 230\,mJ. We argue that
the damage threshold for these masks is, in fact, higher than that and should not be significantly lower than
ablation threshold of fused silica, which is of the order of 1\,J/cm$^{2}$ for femtosecond laser pulses.    
We apply our phase masks for the generation of intense vortex beams and investigate distortions in the vortex beam
patterns that result from several common types of mask defects. 

To the best of our knowledge, the highest previously reported peak intensity of a femtosecond optical vortex beam, 
that has been experimentally generated, is about 200\,GW \cite{vortexPRL}. 
We here report the generation of intense femtosecond vortex beams of various orders with the maximum peak intensity of about 4.6\,TW.
We show that self-focusing of such beams in air results in the generation of  
meter-long ring patterns of multiple ($\sim$30) plasma filaments.

\section{Mask Fabrication}

The fabrication of large-area fused-silica phase masks involves two major steps.
In the first step, the desired grayscale phase pattern is lithographically imprinted into a several micron-thick layer of polyimide
photoresist atop a fused silica substrate, using a Mask-less Lithography Tool (MLT). 
MLT makes use of a unidirectional raster stage and a scanning polygonal mirror which directs 
an externally modulated Argon-Ion Innova Sabre Laser by Coherent, Inc. operating at the wavelength
of 355\,nm and generating output power variable from 100mW to 1.2W. 
The resist exposure level is tailored by adjusting the laser output power as well as the number of
exposure steps and the development time, in order to reach the desired depth in the photoresist.
Phase patterns imprinted into photoresist are profiled with a Veeco NT9800 Optical Profiling System.  

\begin{figure}[b]
\centering\includegraphics[width=8cm]{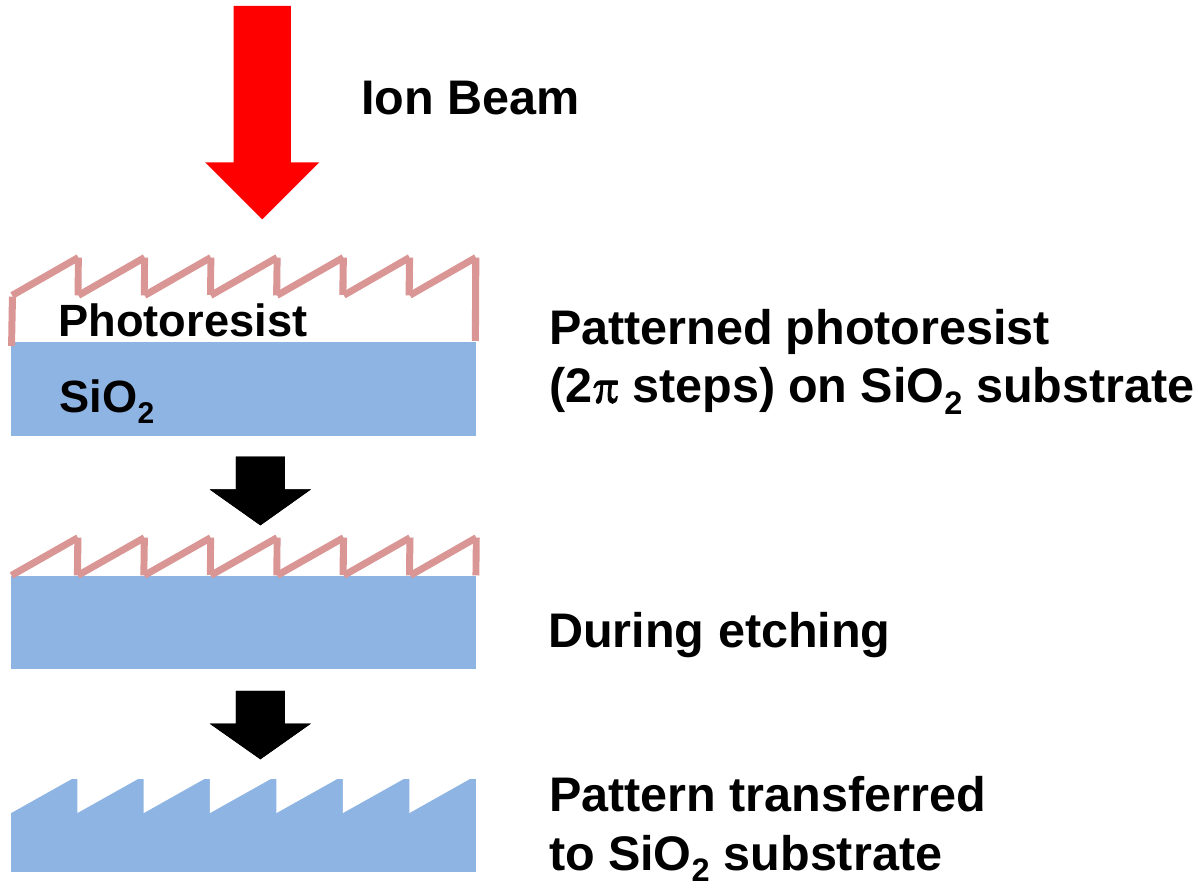}
\caption{Illustration of the second step of the masks fabrication process. The desired grayscale phase
pattern, imprinted into a several micrometer-thin layer of photoresist atop a fused-silica substrate, 
is etched into the substrate via reactive ion etching.} 
\label{fig1}
\end{figure}

In the second fabrication step, illustrated in Figure~1, the pattern is transferred from the polyimide photoresist to the fused
silica substrate via reactive ion etching (RIE). 
The gas mixture used in the RIE process consists of O$_{2}$, Ar, and CHF$_{3}$ gases with flow rates of 
2.8 standard cubic centimeters (sccm), 4 sscm, and 34 sscm, respectively. The RF power used in RIE is 300 watts,
while the pressure in the etch chamber is maintained at 50 mTorr. No adverse effects such as boiling of photoresist
were observed under these etching conditions.  
 
Phase patterns on all masks that we fabricate 
are defined modulo 2$\pi$, meaning that whenever the local phase ramp across the surface of the mask reaches 
the value of 2$\pi$, it is reset to zero. The rates at which RIE removes photoresist and fused silica are
somewhat different. This difference in etching rates depends on the formulation of the resist, as well as on the
exact composition and flow rates of gases used in RIE. In our particular case, the etching rates for resist and fused silica 
differed by about 25\%. Therefore, in order to achieve the target phase contrast on the final fused-silica mask,
the mask on the photoresist had to be made 25\% thicker than the target depth of the mask on fused silica.

\section{Generation of Intense Femtosecond Vortex Laser Beams}

The specific beam shapes that we generated using our fused-silica masks are optical vortices of various orders.
An optical vortex beam is generated from a flat-top or a Gaussian beam through the application of phase-only beam modulation 
in the form:
\begin{eqnarray}
\Phi (r, \theta) \, = \, i m \theta \, ,
\end{eqnarray}
followed by either propagation into far field or focusing the beam with a lens or a curved mirror \cite{vortex}. 
In the above formula, $\Phi$ is a variable phase
that is applied to the transverse amplitude distribution of the beam by the phase mask, $(r , \theta)$ are polar coordinates, 
and $m$ is the order of the vortex. The resulting intensity pattern of the vortex beam, either in far field or in the focal plane of
a focusing optic, looks like a doughnut. The diameter of the intensity doughnut grows approximately linearly 
with the order of vorticity $m$.

\begin{figure}[b]
\centering\includegraphics[width=7cm]{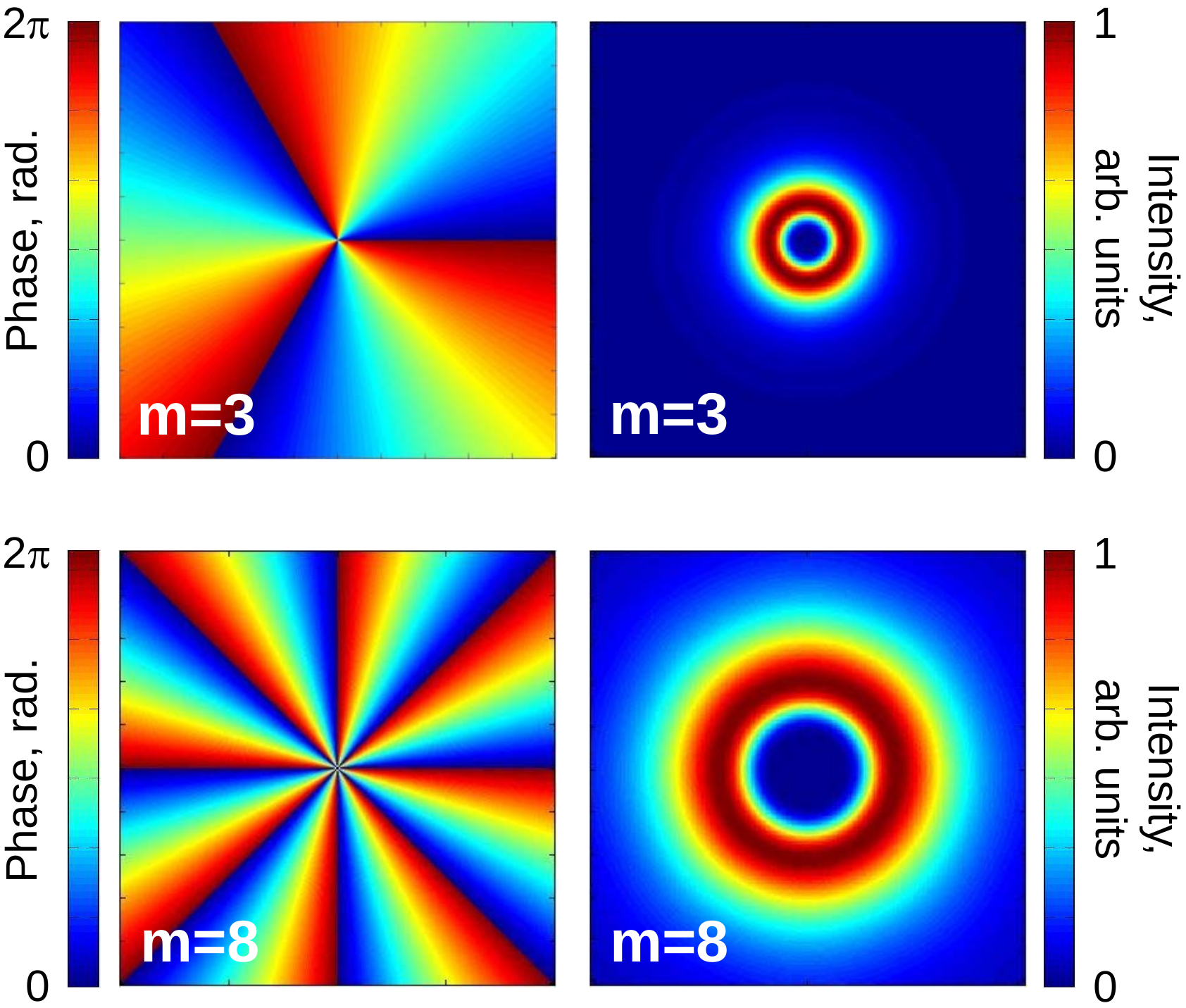}
\caption{Examples of numerically simulated ideal vortex beam patterns of orders 3 and 8. 
The phase profiles, defined modulo 2$\pi$, are shown in the left panels, with the corresponding transverse beam intensity distributions 
shown in the right panels.} 
\label{fig2}
\end{figure}

Although our mask fabrication process allows us to generate vortex beams of arbitrary order, we here limit our consideration to the
cases of vortices of orders 3 and 8. The phase profiles of the corresponding masks, defined modulo 2$\pi$, together with the
corresponding intensity distributions that have been numerically computed and therefore are ideal (free of any defects or imperfections), 
are shown in Figure~2. 
 
In real life, neither the phase masks nor the input laser beam are ideal. Various phase defects that result from the imperfections
inherent to the mask fabrication process, as well as spurious amplitude and phase modulations of the input beam translate into
deviations of the generated vortex beam patterns from their ideal unperturbed shapes such as the ones shown in Figure~2.
Specifically, the peak intensity of the non-ideal vortex ring will be not uniform but will undulate along the circumference of the ring
and the intensity on the beam axis will be different from zero, as would be the case for an ideal optical vortex. For ultra-intense
vortex laser beams, imperfections of the beam pattern may result in the premature azimuthal collapse of the beam and in its
fragmentation into multiple individual filaments that populate the doughnut intensity feature. Such azimuthal mode of collapse 
may precede the self-similar collapse of the intensity ring, as reported in \cite{vortexPRL}. The azimuthal fragmentation of the
vortex intensity ring is not a necessarily undesirable effect in practice, as it results in the deterministic placement of 
intense laser filaments on the ring. In that case, the resulting bottle-like distribution of plasma filaments will not fluctuate 
from one laser shot to another, which, in fact, may be beneficial in some applications. 

\begin{figure}[b]
\centering\includegraphics[width=7cm]{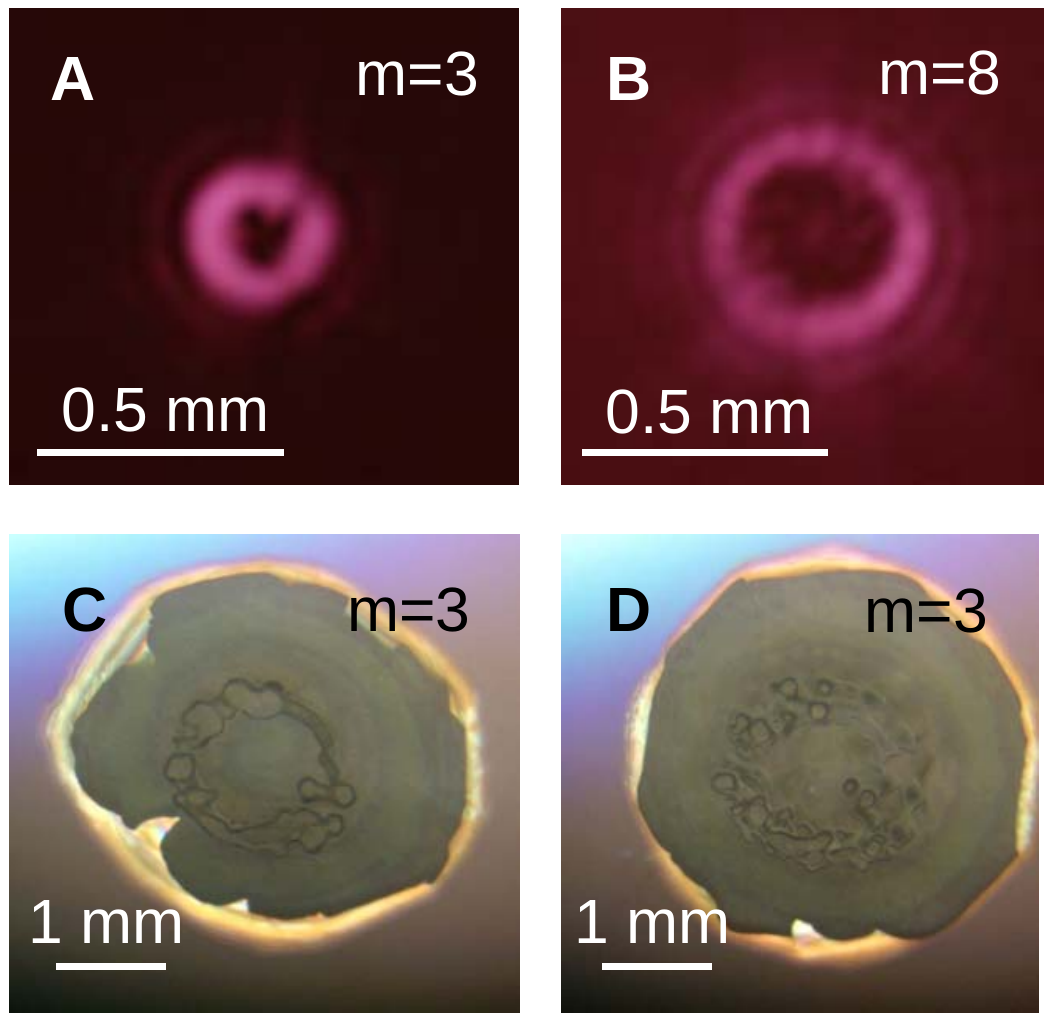}
\caption{(A,B) Intensity distributions of vortex beams of orders 3 and 8 in the linear propagation regime (at low intensity).
The 1/$e ^{2}$ $E$-field radius of the Gaussian beam incident on the masks equals 5\,mm, and the focal length of the focusing lens
used is 1\,m. Distortions of the doughnut-shaped beam intensity patterns are evident. 
(C,D) Bottle-like distributions of plasma filaments produced through self-focusing of an intense optical vortex of order 3 in air.
Focusing conditions are specified in the text. The energy of the 50\,fs-long laser pulse is 150\,mJ (C) and
230\,mJ (D), corresponding to the highest peak laser power of 4.6\,TW.} 
\label{fig3}
\end{figure}

In the top part of Figure~3, we show the intensity patterns of experimentally generated optical vortices of orders 3 and 8.
In order to generate these beam shapes, the input laser beam from a Ti:Sapphire laser oscillator was telescoped to the beam diameter
of 5\,mm, passed through the appropriate phase mask fabricated according to the process described above and  
focused with a lens with a focal length equal to 1 meter. The laser was attenuated to ensure linear propagation through 
the phase masks and focusing optics, as well as through the entire optical path in the air. Thus generated vortex beam patterns were
photographed in the focal place of the lens with a CCD camera sensitive to 800\,nm light, which is the center wavelength of
the laser source. The deviations of the beam patterns from ideal vortex beams are evident for both examples shown.
Specifically, the on-axis intensity is not identically zero, which is especially evident for the case of the 8$^{\mbox{th}}$ order vortex.
Furthermore, the intensity of the doughnut feature is not uniform along its circumference.
The imperfections in the phase masks that may have caused these beam distortions are discussed in detail in the following sections
of the paper.

In the bottom part of Figure~3, we show the ring patterns of multiple plasma filaments that resulted from self-focusing of
an intense femtosecond vortex beam of order 3 in air, at two different levels of optical power. 
The laser source used to produce these filament patterns
is a multi-Terawatt Ti:Sapphire laser system that generates 50\,fs-long pulses at a rate of ten pulses per second. 
At the time when these experiments were conducted,
the maximum attainable compressed pulse energy was about 230\,mJ, corresponding to the peak laser power of about 
4.6 Terawatt. The output beam diameter from the laser is 3\,cm. The beam is converted into an optical vortex of order 3
through the application of a 50\,mm-diameter fused silica vortex phase mask with the thickness of 0.5\,mm, followed by beam focusing 
with a curved mirror with the focal length of 5\,m. No optical damage of the mask was noticed after a continuous extended exposure to 
multiple laser shots, up to the maximum pulse energy of 230\,mJ. At that point the peak fluence and intensity of the laser beam 
incident on the phase mask were 33\,mJ/cm$^{2}$ and 0.65\,TW/cm$^{2}$, respectively. 
There is no apparent reason for the damage threshold of these masks to be
be significantly different from the laser ablation threshold of fused silica, which, for $\sim$50\,fs-long 
laser pulses at 800\,nm wavelength, is of the order of 1\,J/cm$^{2}$ \cite{LLNL}.  

As the intense vortex laser beam propagates in air, beam self-focusing leads to the generation of multiple plasma
filaments that are evident in the images of single-shot burns made by the beam on the front surface of a computer CD, shown in 
the bottom part of Figure~3. The broad and approximately round gray area evident in the images results from the blown-off foil on the back
side of the CD. The number of plasma filaments grows as the laser pulse energy is increased. 
In the two particular cases shown in Figures~3 (A) and (D), the laser pulse energy is 150\,mJ and 230\,mJ, respectively.
The filament patterns shown are imaged at 60\,cm before the focal plane of the focusing optic. Closer to the focal plane,
the filament pattern becomes severely distorted and does not look like a ring anymore. At the highest pulse-energy level of 230\,mJ,
the number of plasma filaments populating the doughnut-shaped intensity feature of the beam is about 30.
      
\section{Distortions of Vortex Patterns Resulting from Common Types of Mask Defects}

In what follows, we will discuss common types of imperfections that occur in the fabrication process of vortex phase masks.
Through numerical simulations, we will quantify errors in the vortex beam patterns that result from these imperfections.
Our simulations assume linear propagation of the beam and are based on the straightforward numerical calculation of the Fresnel diffraction integral.
In all cases shown, the beam incident of the phase mask is assumed to be Gaussian with the 1/$e^{2}$ $E$-field radius of
5\,mm, at 800\,nm wavelength. Immediately following the mask, the beam is focused by a focusing optic with the focal length of 1 meter.

In practice, the acceptable modulation of the vortex intensity pattern will depend on the particular application of the vortex beam.
Our results will set quantitative limits on common-type imperfections of the phase mask, used to generate the beam, 
at a given maximum acceptable beam distortion.  

\subsection{Error in 2$\pi$ Phase Step}

\begin{figure}[b]
\centering\includegraphics[width=8.5cm]{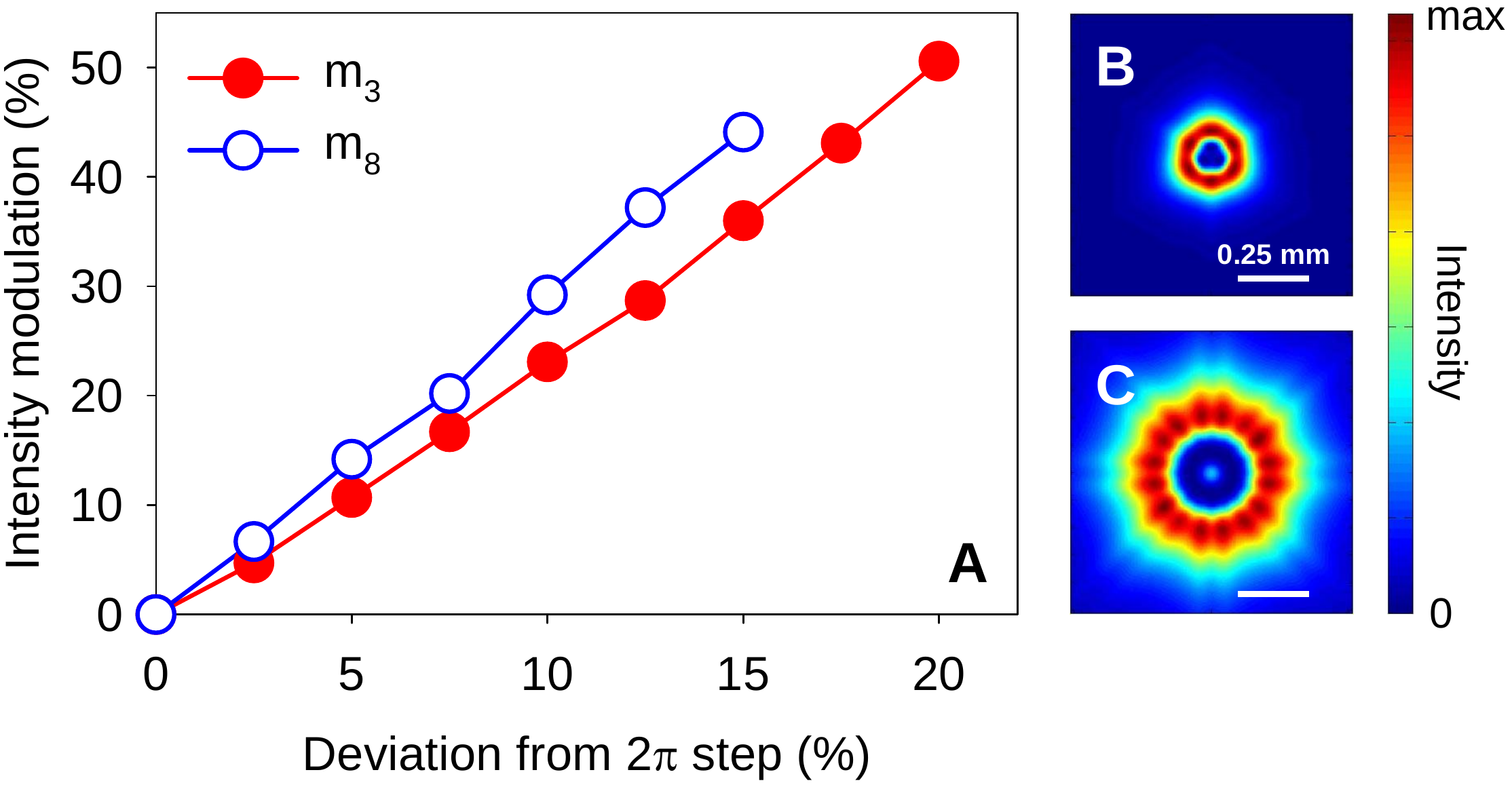}
\caption{(A) Peak-to-peak intensity modulation of the doughnut intensity feature of the beam as a function
of the \% error in the maximum phase step of the phase mask, for continuous-wave optical vortices of orders 3 and 8.
(B) Beam intensity pattern in the focal plane of the focusing optic, for an optical vortex of order 3 and for 5\% error in the 
maximum phase step. (C) Same for a vortex of order 8.} 
\label{fig4}
\end{figure}

As we pointed out above, phase patterns on all phase masks that we fabricate are defined modulo 2$\pi$. Whenever the phase ramp across 
the surface of the mask reaches 2$\pi$, the phase is reset to zero. Phase modulation imposed on the beam by the mask, at a particular 
point on the mask, is proportional to the mask thickness at that point and inversely proportional to the beam wavelength. Therefore,
if the thickness contrast on the fabricated mask deviates from the target value, the value of the maximum phase modulation by the mask 
will be different from 2$\pi$. The same effect will result if the mask is used to modulate a laser beam having a different wavelength 
from the one it is designed for.

In Figure~4 (B,C), we show two examples of imperfect vortex beam patterns that result from the deviation of the maximum
phase contrast by 5\% from the target value of 2$\pi$. In the Figure~4(A), we show the resulting azimuthal modulation of the
ring intensity pattern, as a function of the \% error in the maximum phase contrast of the phase mask used to generate the beam.
The intensity modulation of the ring is defined as the maximum relative peak-to-peak variation of intensity along the circumference 
of the intensity doughnut. As evident from these simulations, the modulation of the ring intensity pattern grows approximately linearly 
with the \% error in the maximum phase contrast of the phase mask. The beam distortion is somewhat higher for larger vortex orders.
In addition to the modulation of the ring intensity feature, imperfect higher-order vortices exhibit pronounced on-axis intensity.

\subsection{Effect of Broad Spectrum of Light}

\begin{figure}[b]
\centering\includegraphics[width=8.5cm]{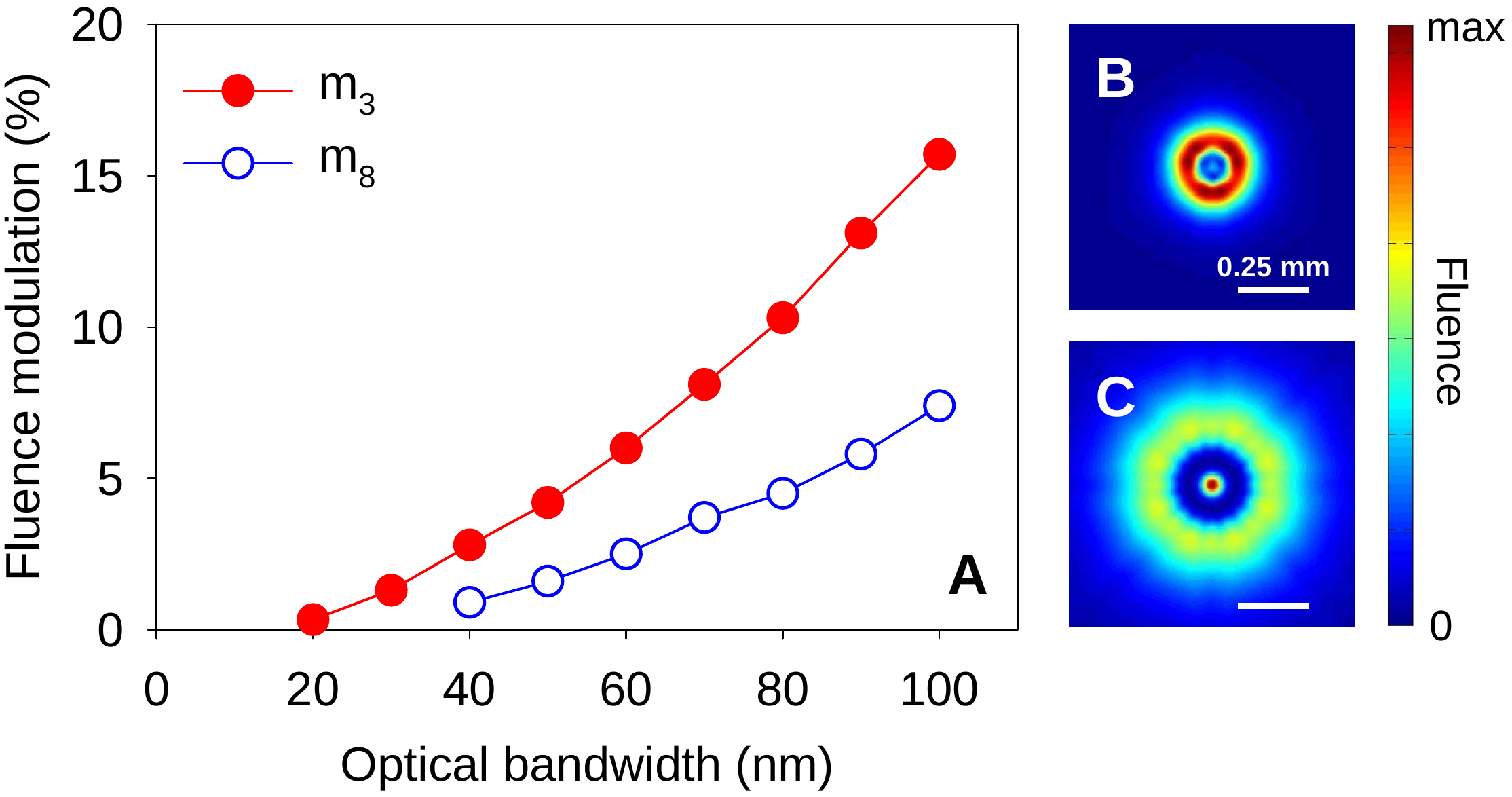}
\caption{(A) Peak-to-peak fluence modulation of the vortex ring as a function of the optical bandwidth of the laser beam.
(B) Fluence pattern for an optical vortex of order 3 and for optical bandwidth equal to 100\,nm. (C) Same for a vortex of order 8.} 
\label{fig5}
\end{figure}

The second type of distortion of optical vortex beams that we consider results from the 
finite optical bandwidth of the incident laser beam. Even a perfectly fabricated mask will be ideal only for one optical wavelength
in the spectrum of the laser. In the calculation of the beam distortion in this case, we compute the spatial distribution
of the beam fluence (intensity integrated over pulse duration), in the focal plane of the focusing optic.

The result of the calculation is shown in Figure~5. As evident from the data, the ring modulation for higher-order vortices is, in fact,
smaller than that for lower-order vortices. This trend results from averaging of errors for different spectral components of the beam 
that becomes more effective as the order of vorticity is increased. We point out that for higher-order vortices, the emergence of significant 
intensity on the beam axis becomes the dominant type of error for the case of pulsed laser beams with large optical bandwidth.

The typical value of optical bandwidth for amplified Ti:Sapphire laser chains is about 30\,nm, in which case the modulation 
of the vortex ring patterns caused by the finite bandwidth of the laser source is of the order of one percent or lower.  

\subsection{Circular Defect}

\begin{figure}[b]
\centering\includegraphics[width=7cm]{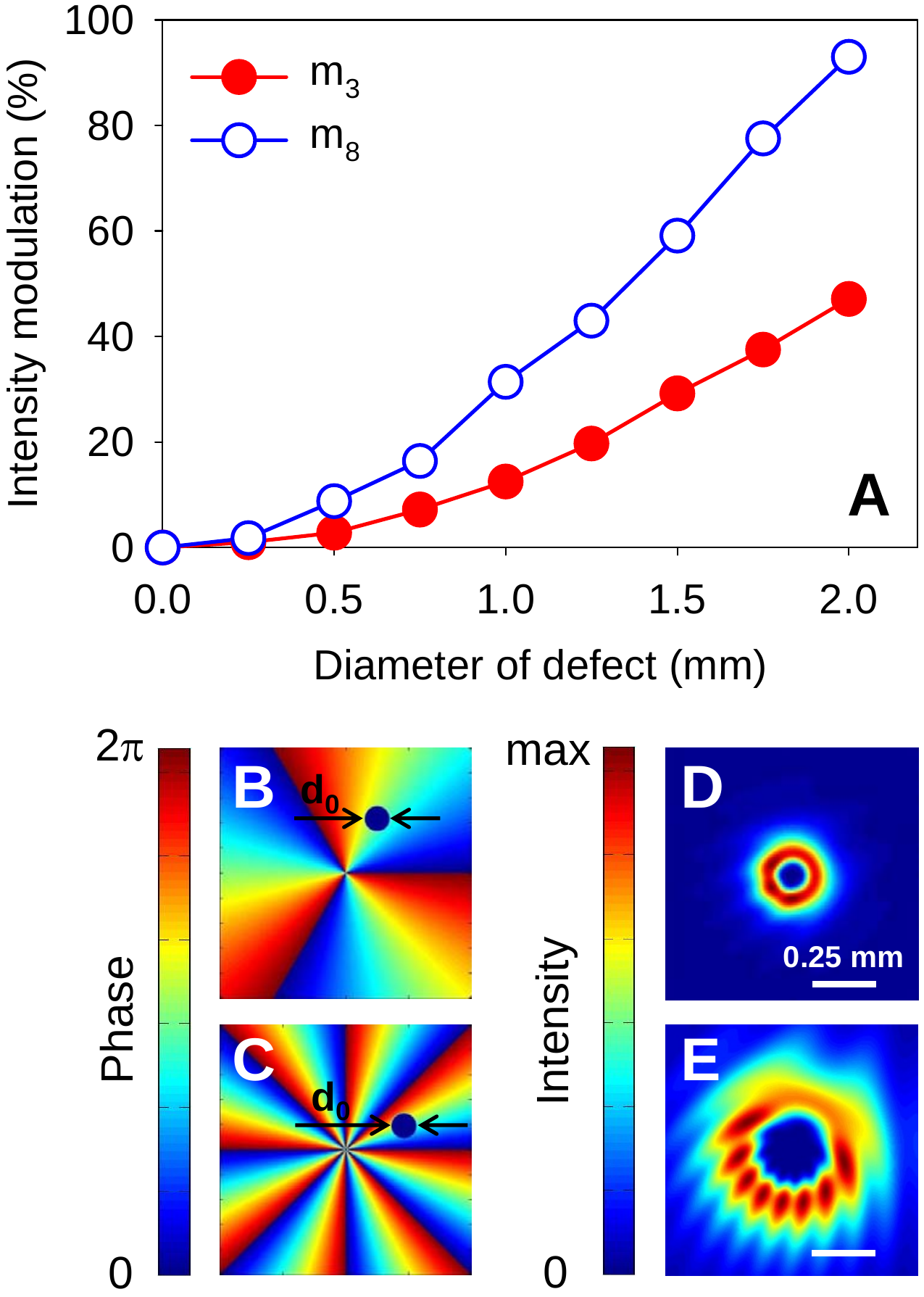}
\caption{(A) Peak-to-peak intensity modulation of the vortex ring as a function of the diameter of the circular defect on the phase mask.
(B,C) Illustrations of the location of the circular defect on vortex phase masks of orders 3 and 8. (D,E) Corresponding beam patterns
for the particular diameter of the circular defect equal to 1\,mm.} 
\label{fig6}
\end{figure}

Another type of defect that is common in mask fabrication is a point or circular defect. The effect of this type of defect on the 
generated vortex beam pattern will depend on the location and size of the defect. To get an idea about severity 
of the vortex beam distortion resulting from this type of defect, we assume that the defect is located at a median of the 2$\pi$ 
phase ramp on the mask, and its center is one 1/$e^{2}$ beam radius away from the center of the mask. We model the defect
as uniform zero phase inside a circle of a particular diameter $d_{0}$, as illustrated in Figures~6(B,C) for vortex masks of
orders 3 and 8, respectively. The calculation in this case is performed for an otherwise ideal phase mask and for a continuous-wave laser beam.

As follows from the results of the calculation, the modulation of the ring intensity pattern grows approximately quadratically with 
the diameter of the defect, which corresponds to the approximately linear growth with the area of the defect. For the same defect size, 
beam distortion is more severe for larger vortex orders.    

\subsection{Line Defect}

\begin{figure}[bt]
\centering\includegraphics[width=7cm]{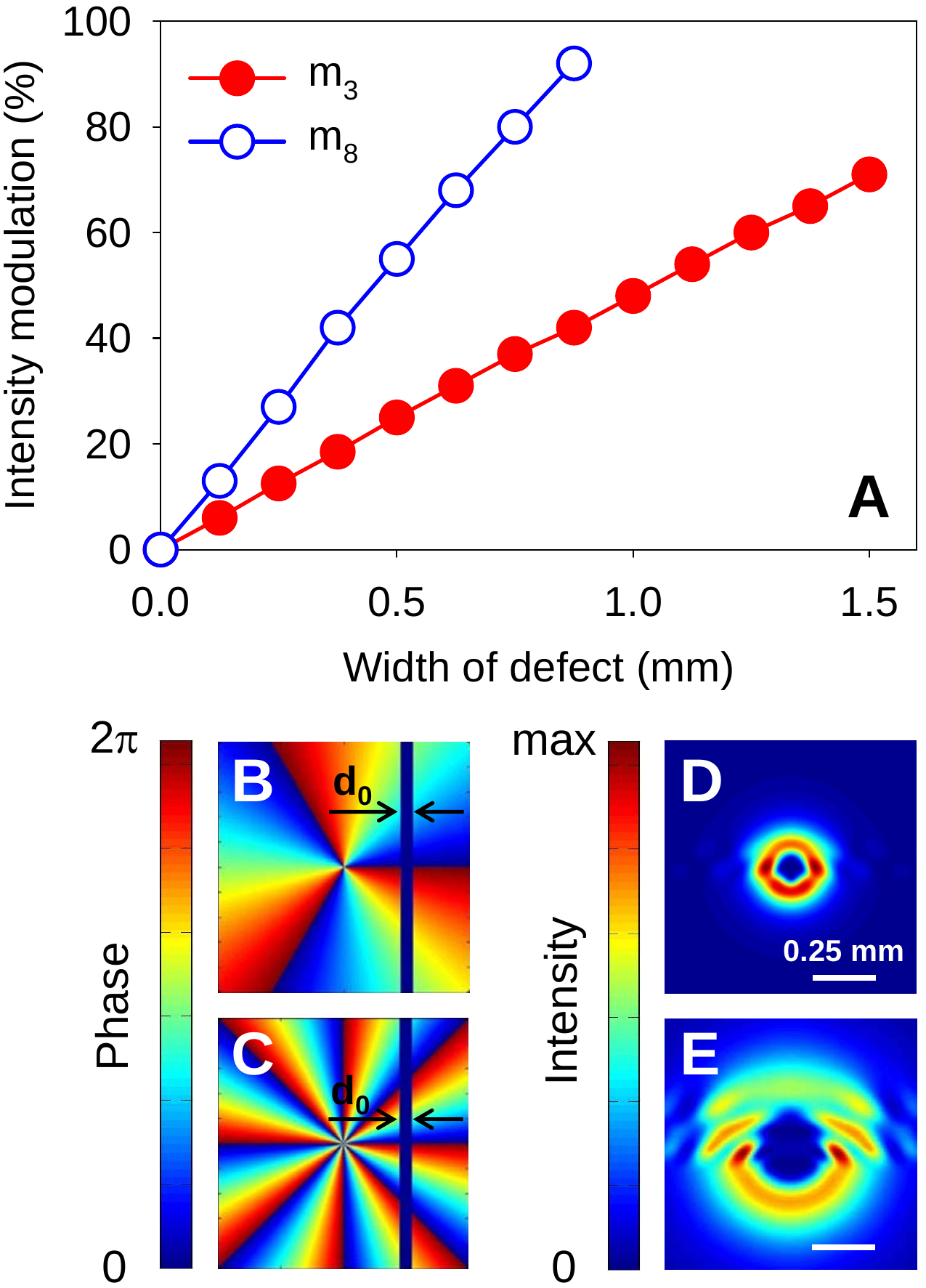}
\caption{(A) Peak-to-peak intensity modulation of the vortex ring as a function of the width of the line defect on the phase mask.
(B,C) Illustrations of the location of the linear defect on vortex phase masks of orders 3 and 8. (D,E) Corresponding beam patterns
for the particular width of the linear defect equal to 0.5\,mm.} 
\label{fig7}
\end{figure}

The last defect type we consider is the line defect. This defect type will result, for example, from stitching of several 
independently fabricated phase masks into a single large-area mask.
As in the case of the circular defect, the appearance and severity of
beam distortion in this case will depend on the size of the defect and on its location on the mask. To be specific, we assume that the 
defect is perpendicular to one of the 2$\pi$ phase-reset lines and is located one 1/$e^{2}$ beam radius away from the center of the
mask, as illustrated in Figures~7(B,C), for vortex orders 3 and 8. As in the case of a circular defect, we model line defect as uniform
zero phase inside a rectangular area of a particular width $d_{0}$ spanning the entire length of the mask.

The results of the calculation for this case are shown in Figure~7. Intensity modulation of the doughnut intensity feature 
grows approximately linearly with the width of the line defect, and the effect of this type of defect is more
severe for vortices of higher order. 

\section{Conclusion}
In conclusion, we have reported experimental results on the fabrication and application of large-area, grayscale phase masks 
suitable for beam shaping of femtosecond laser systems with multi-Terawatt peak power. 
We have applied these masks for the generation of intense optical vortices of various orders, with peak power of up to 4.6\,TW.
The corresponding maximum peak intensity of the laser beam incident on the surface of the masks 
was 0.65\,TW/cm$^{2}$. The pulse energy of optical vortex beams generated in our experiments is by more than one order of magnitude higher 
than what was reported previously. The peak intensity of vortex beams in our experiments was sufficient for the generation 
of meter-long bottle-shaped patterns of multiple plasma filaments in ambient air.
Through numerical simulations, we have analyzed distortions of the generated vortex beam patterns that result from common types of 
mask defects. Our simulation results will be useful for estimations of maximum acceptable mask imperfections for a given
level of modulation of vortex beam patterns.     

\,
\vspace{0.5cm}

This work was supported by The United States Air Force Office of Scientific Research (U.S. AFOSR) under Grants 
No. FA9550-12-1-0143, No. FA9550-12-1-0482, and No. FA9550-10-1-0561 and by The United States Defense Threat
Reduction Agency (DTRA) under Grant No. HDTRA1-14-1-0009. Craig Ament acknowledges the support from
the Graduate Scholarship by the Directed Energy Professional Society (DEPS).

\end{document}